\documentclass[11pt,peerreview,onecolumn,doublespace]{IEEEtran}

\title{Transmit Power Minimization for MIMO Systems of Exponential Average BER with
Fixed Outage Probability
\thanks{This work was supported in part by the Scheme of Research Exchanges with China and India,
the Royal Academy of Engineering of UK,  the Specialized Research
Fund for the Doctoral Program of Higher Education under Grant
20132125110006, and the Fundamental Research Funds for the Central
Universities under grant 3132013334.}
\thanks{Dian-Wu Yue is with the College of Information Science and
Technology, Dalian Maritime University, Dalian, Liaoning 116026,
China (e-mail: dianwuyue@yahoo.com). Yichuang Sun is with the
School of Engineering and Technology, University of Hertfordshire,
College Lane, Hatfield, Herts AL10 9AB, UK, (Email:
y.sun@herts.ac.uk). } }

\author{Dian-Wu Yue  and  Yichuang Sun}

\usepackage{graphicx,amssymb,amsmath}
\usepackage{multicol}
\usepackage{anysize}
\usepackage{color}
\usepackage{cite}
\usepackage{balance}

\newcommand{\be}{\begin{equation}}
\newcommand{\ee}{\end{equation}}
\newcommand{\bee}{\begin{eqnarray}}
\newcommand{\eee}{\end{eqnarray}}
\newcommand{\nnb}{\nonumber}

\newtheorem{Lemma}{Lemma}

\newtheorem{Proposition}{Proposition}

\begin{document}

\maketitle

\begin{abstract}
This paper is concerned with a wireless system operating in MIMO
fading channels with channel state information being known at both
transmitter and receiver.  By {\it spatiotemporal} subchannel
selection and power control,  it aims to minimize the average
transmit power (ATP) of the MIMO system while achieving an
exponential type of average bit error rate (BER) for each data
stream. Under the constraints of a given fixed individual outage
probability (OP) and average BER for each subchannel, based on a
traditional upper bound and a dynamic upper bound of Q function,
two closed-form ATP expressions are derived, respectively, and
they correspond to two different power allocation schemes.
Numerical results are provided to validate the theoretical
analysis, and show that the power allocation scheme with the
dynamic upper bound can achieve more power savings than the one
with the traditional upper bound.

\end{abstract}


\newpage

\section{Introduction}

Wireless transmission using multiple antennas has attracted much
interest in the past couple of decades due to its capability to
exploit the tremendous capacity inherent in multiple-input
multiple-output (MIMO) channels \cite{Telatar}. Various aspects of
wireless MIMO systems have been studied intensively. All
theoretical analysis for MIMO systems in the literature can be
roughly divided into two categories: capacity analysis for the
system efficiency \cite{Kiessling} and performance analysis for
the system reliability \cite{Jin}. Certainly, there are some
papers in between that simultaneously consider the system
efficiency and reliability, leading to some fundamental tradeoff
between the two \cite{Zheng}-\cite{Falou}.

Adaptive transmission techniques can utilize the resources
efficiently and thus, are always of great interest in the field of
wireless communications, especially for the current multiple
antenna systems. The basic motivation behind adaptive transmission
is to obtain improvements in terms of average spectral efficiency
or bit-error rate (BER) by exploiting the channel knowledge
available at transmitter. The optimal power control policy that
maximizes the fading channel capacity is shown to be of
waterfilling type from the information theoretic point of
view\cite{Caire}. On the other hand, dating back to early 1968,
Hayes \cite{Hayes} considered adaptive power control problem for
single antenna systems, resulting in an optimal power strategy
that minimizes the BER subject to an average power constraint.
Based on optimization theory and random matrix theory, several
novel adaptation transmit schemes including optimal power control
policies in multiple antenna systems have been already proposed
\cite{Palomar}-\cite{Zanella2}.

It is well known that in an additive white Gaussian noise (AWGN)
channel, a single-input single-output (SISO) wireless system with
coherent signalling schemes \cite{Xiong}  can have a BER
exponentially decreasing with the signal-to-noise ratio (SNR), or
equivalently, achieve an infinite diversity order.  When the same
system operates in a Rayleigh fading channel, however, its average
BER decreases only inversely with the SNR. The degradation can be
partially mitigated if we replace the SISO system with MIMO. In
spite of various efforts,  nearly all existing MIMO system schemes
can only achieve a finite diversity order, even with spatial power
control \cite{Paulraj},\cite{Zhou}.

However, Rangarajan {\em et al} showed for the first time in
\cite{Rangarajan1} and \cite{Rangarajan2} that by adaptive power
control in time, a SISO system can have a BER performance with
exponential diversity order in Rayleigh fading. Subsequently,
Sharma {\em et al} analyzed optimal adaptive power transmission
policies for MIMO systems, and showed how to use a combined
temporal and spatial adaptive policy to obtain an exponential
diversity order for MIMO systems in Rayleigh fading \cite{Sharma},
\cite{Premkumar}. The aforementioned results require perfect
channel-state information (CSI) at both the transmitter and
receiver. In \cite{Khan1}, Khan further showed that the
exponential diversity order can be achieved in ``all'' fading
channels. Moreover, the authors in \cite{Muller} presented two
different power allocation strategies of achievable exponential
diversity order for wireless multihop systems. In addition to the
total average power constraint, when the more realistic scenario
of peak to average power ratio (PAPR) constraint is also
satisfied, papers \cite{Ganesh} and \cite{Reddy} considered such
an optimal power control problem for MISO channels and obtained
the minimized BER of exponential diversity. Even in the practical
case with imperfect CSI at the transmitter, papers \cite{Khan2}
and \cite{Khan3} showed that in Rayleigh fading environments the
exponential diversity order can also be obtained by appropriate
spatiotemporal power allocation.

It should be noticed that in the existing techniques to achieve an
exponential diversity order as mentioned above, an MIMO system is
only allowed to transmit a single information stream along one of
its eigen beams. Although orthogonal space-time block coding
(OSTBC) is also discussed in \cite{Sharma}, it is indeed
equivalent to a SISO (or alternatively, called single beamforming)
system \cite{Larrson}. In this paper, we will adopt multi-channel
beamforming (\cite{Jin},\cite{Palomar}) to utilize efficiently the
degree of freedom provided by multiple antennas. In addition,
obviously different from the optimal control strategy adopted in
\cite{Rangarajan1}-\cite{Khan3} that minimize the system BER under
the average transmit power constraint, we will pursue another
optimal control strategy of minimizing average transmit power
under individual average BER and outage probability (OP)
constraints for each data stream. This strategy is consistent with
the current efforts of green communications \cite{Li},
\cite{Feng}. And the average BER will be expressed as an
exponential function of SNR, which implies that the underlying
MIMO system has exponential diversity order.

The rest of the paper is organized as follows. In Section II, we
describe the system model and present the optimization problems.
In Section III, with the help of an order statistical result of
eigenvalues of complex central Wishart matrices, we derive a
closed-form ATP expression based on the traditional upper bound of
Q function. In Section IV, we present a dynamic upper bound of Q
function, and based on it derive further another closed-form ATP
expression. After that, in Section V we provide some numerical
results to validate the theoretical analysis and make comparisons
between the two different power allocation schemes. Finally, in
Section VI we conclude the paper.

\section{System Model And Optimization Formulation}

\subsection{System model}

We first consider a single-user MIMO system operating in flat
fading environments with $n_T \geq 1$ transmit antennas and $n_R
\geq 1$ receive antennas, and assume that perfect CSI is available
at both the transmitter and the receiver. We denote by $h_{ij}$
the channel gain between $i$-th receive antenna and $j$-th
transmit antenna, and by $\mathbf{H}$ the corresponding channel
gain matrix whose $i,j$-th entry is $h_{ij}$. In Rayleigh fading
environments, it is further assumed that $h_{ij}$ is independent
and identically distributed (i.i.d.) and thus $\mathbf{H}$ follows
the joint complex Gaussian distribution with zero mean matrix and
covariance matrix $\mathbf{I}_{n_R}\otimes \mathbf{I}_{n_T}$,
i.e., $ \mathbf{H} \sim
\mathbb{CN}(\mathbf{\mbox{0}},\mathbf{I}_{n_R}\otimes
\mathbf{I}_{n_T})$ (see \cite{Gupta} for these notations). For a
transmission through the MIMO channel with $\mathbf{H}$, the $n_R
\times 1$ received vector can be expressed as \be \mathbf{y}
=\mathbf{H}\mathbf{x}+\mathbf{n} \ee where $\mathbf{x}$ is the
$n_T \times 1$ transmitted vector and $\mathbf{n}$ is the $n_R
\times 1$ additive noise vector following complex Gaussian
distribution of zero-mean vector and covariance matrix
$\mathbf{I}_{n_R}$, i.e., $\mathbf{n} \sim
\mathbb{CN}(\mathbf{\mbox{0}},\mathbf{I}_{n_R})$.

Now let $m=\min \{n_T,n_R\}$ and $n=\max \{n_T,n_R\}$. Define
\be
    \mathbf{\Omega} =\left\{\begin{array}{ll}
     \mathbf{H}^{\dag}\mathbf{H}, & \mbox{for}\; m=n_T;\\
     \mathbf{H}\mathbf{H}^{\dag}, & \mbox{for}\; n=n_T.  \end{array}
     \right.
 \ee
From Chapter 3 of \cite{Gupta}, it follows that the matrix
$\mathbf{\Omega}$ follows Wishart distribution, i.e.,
$\mathbf{\Omega} \sim \mathbb{CW}(n,\mathbf{I}_m)$. Following the
conventional spatial multiplexing method based on singular value
decomposition(SVD) \cite{Raleigh}, \cite{Lebrun}, the channel
matrix can be written as \be \label{mmm}
\mathbf{H}=\mathbf{U}\mathbf{\Lambda}\mathbf{V}^{\dag} \ee where
$\mathbf{U}$ and $\mathbf{V}$ are unitary matrices, and \be
\mathbf{\Lambda}=\mbox{diag}(\sqrt{\lambda_1},\sqrt{\lambda_2},\ldots,\sqrt{\lambda_m})\ee
with $\{ \lambda_i: i=1,2,\ldots,m \}$ being the eigenvalues of
$\mathbf{\Omega}$ sorted in descending order, i.e., \be
\lambda_1\geq \lambda_2\geq \cdots \geq \lambda_m.    \ee Thus we
can transmit $r \leq m$ data symbols at one time. Since
$\mathbf{H}$ is known perfectly at the transmitter, we can set the
transmitted vector as \be
\mathbf{x}=\mathbf{V}_1^r\mathbf{P}\mathbf{s}   \ee where
$\mathbf{s}$ is the $r \times 1$ modulated data vector with
covariance matrix $\mathbf{I}_{r}$, $\mathbf{V}_1^r$ is the
precoding matrix formed with the first $r$ columns of $\mathbf{V}$
associated with the first $r$ largest eigenvalues of
$\mathbf{\Omega}$,  and $\mathbf{P}$ is a diagonal matrix as
follows: \be
\mathbf{P}=\mbox{diag}(\sqrt{p_1},\sqrt{p_2},\ldots,\sqrt{p_r})\ee
where $\{p_i:i=1,2,\ldots,r\}$ are the powers allocated to the $r$
established data streams. Due to the assumption that CSI is
available at the receiver, the symbols transmitted through the
receive filter are recovered from the received vector $\mathbf{y}$
with matrix $\mathbf{U}_1^r$, defined similarly to
$\mathbf{V}_1^r$, as \bee \hat{\mathbf{s}}
&=&(\mathbf{U}_1^r)^{\dag}(\mathbf{H}\mathbf{x}+\mathbf{n})\nnb \\
&=& \mathbf{\Lambda}_1^r \mathbf{P}
\mathbf{s}+(\mathbf{U}_1^r)^{\dag}\mathbf{n} \eee where
$\mathbf{\Lambda}_1^r$ is a diagonal submatrix of
$\mathbf{\Lambda}$ that contains the $r$ largest eigenvalues in
descending order, and the filter-processed noise
$\mathbf{\eta}=(\mathbf{U}_1^r)^{\dag}\mathbf{n}$ has the same
statistical properties as $\mathbf{n}$, possibly with a reduced
dimension. Each data stream then experiences an instantaneous SNR
given by \be \label{SS-1} \mbox{SNR}_i=\lambda_i p_i,\;\;
i=1,2,\ldots, r. \ee  And the corresponding short term BER is
expressed as \be P_b^{(i)} =\xi_i Q(\sqrt{\beta_i\mbox{SNR}_i})
\ee where $Q(\cdot)$ is the Gaussian $Q$ function, and the
parameters $\xi_i$ and $\beta_i$ are constants, depending on the
used modulation type \cite{Goldsmith2}.

\subsection{Optimization formulation}

 It is well-known that the OP is defined as
the probability when the instantaneous SNR falls below a certain
threshold \cite{Wang}. At this time when the $i$-th subchannel is
in bad condition, in order to save transmit power, the subchannel
should have a transmit outage temporarily.  For this reason, in
order to analyze conveniently, here we set the SNR threshold as $
p_i\overline{\lambda}_{\mbox{\small out}}(i)$ for the $i$-th
subchannel. So we will introduce a transmit outage when $\lambda_i
<\overline{\lambda}_{\mbox{\small out}}(i)$. Accordingly, the
individual OP is expressed as  \be
P_{\mbox{out}}^{(i)}=\int^{\overline{\lambda}_{\mbox{\small
out}}(i)}_0 f_i(\lambda_i)d\lambda_i \ee where $f_i(\lambda_i)$ is
the p.d.f. of eigenvalue $\lambda_i$.

Once the OP is given, we can carry on adaptive transmission. In
particular, based on channel eigenvalues, we can select those MIMO
subchannels satisfying the OP constraint condition to transmit
data streams, and let each of them transmit a data stream. In
order to utilize efficiently MIMO subchannels, we should employ
all those satisfactory subchannels to communicate. Note that if
any subchannel does not satisfy the constraint, then this implies
that the subchannel cannot transmit a data stream, and thereby we
force the channel into the state of channel outage; and if none of
the MIMO subchannels satisfies the constraint, then this will
result in a system outage.

The above-mentioned adaptive transmission involves not only
channel selection but also power control, both of which are
conducted based on the status of eigenvalues of channel matrix. As
already mentioned before, our adaptive power allocation strategy
aims at minimizing the total ATP while each data stream achieves
an exponential average BER. For this reason, under the constraint
that both the individual OP $P_{\mbox{out}}^{(i)} $ and the
individual average BER $\overline{P}_b(i)$ are given, this optimal
problem can be formulated as \be \label{optimal-p}
    \left\{\begin{array}{ll}
      \underset{\{p_i: 1 \leq i \leq m\}}{\mbox{Minimize}} & \rho=\mathbb{E}\{\sum_{i=1}^m p_i\} ;\\
     \mbox{Subject to} & \frac{\mathbb{E}P_b^{(i)}}{\overline{P}_o(i)}
     \leq \overline{P}_b(i),\;\; 1\leq i\leq m
      \end{array}
     \right.
 \ee
where $\overline{P}_o(i)=1-P_{\mbox{out}}^{(i)}$ denotes the
transmit probability, and
$\mathbb{E}P_b^{(i)}=\int_{\overline{\lambda}_{\mbox{\small
out}}(i)}^{\infty}P_b^{(i)}f_i(\lambda_i)d\lambda_i=\int_{\overline{\lambda}_{\mbox{\small
out}}(i)}^{\infty}\xi_i Q(\sqrt{\beta_i\lambda_i
p_i})f_i(\lambda_i)d\lambda_i$ with $\mathbb{E}(\cdot)$ standing
for the expectation operator. On the other hand,  the required BER
$\overline{P}_b(i)$ can be expressed as an exponential function of
SNR: \be \label{Pb-1}
\overline{P}_b(i)=\frac{\xi_i}{2}e^{-\beta_i\widehat{\mbox{SNR}}(i)/2}.
\ee It should be pointed out that the SNR
$\widehat{\mbox{SNR}}(i)$ can be designed beforehand.

Obviously, this optimization problem can be translated into $m$
individual optimization problems, and each corresponds to an
ordered subchannel: \be
    \left\{\begin{array}{ll}
      \underset{\{p_i\}}{\mbox{Minimize}} & \mathbb{E}\{p_i\} ;\\
     \mbox{Subject to} &\mathbb{E}P_b^{(i)}
     \leq \overline{P}_o(i)\overline{P}_b(i).
      \end{array}
     \right.\ee
Applying Lagrange Multiplier Method to each of the above
sub-optimization problems, we get the following family of
unconstrained optimization problems parameterized by multipliers
$\omega_i>0$, $\;\; 1 \leq i \leq m $: \be
\underset{\{p_i\}}{\mbox{Min}}\int_{\overline{\lambda}_{\mbox{\small
out}}(i)}^{\infty}p_if_i(\lambda_i)d\lambda_i+\omega_i\int_{\overline{\lambda}_{\mbox{\small
out}}(i)}^{\infty}P_b^{(i)}f_i(\lambda_i)d\lambda_i-\omega_i\overline{P}_o(i)\overline{P}_b(i).\ee
or \be
\underset{\{p_i\}}{\mbox{Min}}\int_{\overline{\lambda}_{\mbox{\small
out}}(i)}^{\infty}[p_i+\omega_i(P_b^{(i)}-\overline{P}_b(i))]f_i(\lambda_i)d\lambda_i.\ee
If we make use of the exact expression of $P_b^{(i)} =\xi_i
Q(\sqrt{\beta_i\mbox{SNR}_i})$ to solve the problems, then due to
the relatively complicated Q function, we can only have an
unclosed-form expression based on the Lambert W function
\cite{Lee}. Similar to \cite{Sharma} and \cite{Muller},  we also
employ the common upper bound $Q(x) \leq \frac{1}{2}e^{-x^2/2}$ to
replace the exact expression and obtain easily a suboptimum
solution as follows: \be  \label{pp-2}
p_i=\left\{\begin{array}{ll}
\frac{2}{\beta_i\lambda_i}\ln(\frac{\lambda_i}{\lambda_0^{(i)}}) & \mbox{for} \lambda_i > \lambda_0(i);\\
0 & \mbox{for} \lambda_i \leq \lambda_0(i).
     \end{array}
     \right.\ee
where $\lambda_0(i)=\max
\{\lambda_0^{(i)},\overline{\lambda}_{\mbox{\small out}}(i)\}$ and
$\lambda_0^{(i)}$ can be found by solving \be
\int_{\lambda_0(i)}^{\infty}
\frac{\xi_i}{2}e^{-\beta_ip_i\lambda_i/2}f_i(\lambda_i)d\lambda_i=\overline{P}_o(i)\overline{P}_b(i).\ee
This suboptimum solution will provide convenience for us to
produce theoretical and numerical results.

\section{Minimum Average Transmit Power and A Power Allocation Scheme}

\subsection{Individual outage probability}

It follows from \cite{Zanella} that the marginal p.d.f. of the
$i$-th largest eigenvalue $\lambda_i, \; i=1,2,\ldots,m$, can be
expressed as a sum of terms $\lambda_i^a e^{-b\lambda_i}$, which
is very friendly for further analysis. By the expression, we can
easily get the following expression of individual outage
probability.

\begin{Lemma} The individual OP for the $i$ data stream can be
given by  \be  P_{\mbox{out}}^{(i)}=\sum_{k=i}^m(-1)^{k-i}{k-1
\choose i-1}{m \choose
k}\overline{F}_{\mbox{min:}k}^{\mbox{(out)}}(\overline{\lambda}_{\mbox{\small
out}}(i))\ee where
$\overline{F}_{\mbox{min:}k}^{\mbox{(out)}}(\overline{\lambda}_{\mbox{\small
out}}(i))$ denotes the distribution function of the smallest
random variable considered in a subset of $k$ random variables
over the random variable set of all eigenvalues $\{\lambda_i, \;
i=1,2,\ldots,m \}$, and is given by \cite{Zanella} \bee
&&\overline{F}_{\mbox{min:}k}^{\mbox{(out)}}(\overline{\lambda}_{\mbox{\small
out}}(i))=\frac{kC}{m!}\sum_{\mathcal{\alpha}}\sum_{\mathcal{\mu}}
\mbox{sgn}(\mathcal{\alpha})\mbox{sgn}(\mathcal{\mu})A_{k}(\mathcal{\alpha},\mathcal{\mu})\nnb \\
&& \times \mathbb{\sum}_{\mathcal{\tau}}
\frac{\gamma(\theta+\alpha_k+\mu_k-1+\sum_{\iota=1}^{k-1}\tau_{\iota},k\overline{\lambda}_{\mbox{\small
out}}(i))}{k^{\theta+\alpha_k+\mu_k-1+\sum_{\iota=1}^{k-1}\tau_{\iota}}}
\nnb \\
&&\times\prod_{\iota=1}^{k-1}\frac{(\theta+\alpha_{\iota}+\mu_{\iota}-2)!}{\tau_{\iota}!}
\eee where $\theta=n-m$, $C$ is a constant standing for \be C=
\frac{1}{\prod_{j=1}^m(m-j)!\prod_{j=1}^m(n-j)!}, \ee
$\mbox{sgn}(\mathcal{\alpha})$ denotes the sign of permutation
$\mathcal{\alpha}=(\alpha_1,\alpha_2,\ldots,\alpha_m)$ for
integers $\{1,2,\ldots,m\}$,
$A_{k}(\mathcal{\alpha},\mathcal{\mu})$ is defined as \be
A_{k}(\mathcal{\alpha},\mathcal{\mu})=\prod_{\iota=k+1}^m
(\theta+\alpha_{\iota}+\mu_{\iota}-2)! ,\ee and
$\mathbb{\sum}_{\mathcal{\tau}}$ denotes \be
\mathbb{\sum}_{\mathcal{\tau}}=\sum_{\tau_1=0}^{\theta+\alpha_1+\mu_1-2}\sum_{\tau_2=0}^{\theta+\alpha_2+\mu_2-2}
\cdots\sum_{\tau_{k-1}=0}^{\theta+\alpha_{k-1}+\mu_{k-1}-2}.\ee
Moreover, $\gamma (q,x)$ is just the incomplete gamma function
(See Page 454 of \cite{Andrews}).
\end{Lemma}
On the other hand, the global outage probability for the whole
system is written as \bee P_{\mbox{out}}&=& \mbox{Prob}(\lambda_i
<\overline{\lambda}_{\mbox{\small out}}(i),\;\; 1\leq i \leq m)\nnb \\
&\leq & P_{\mbox{out}}^{(1)}.\eee When
$\overline{\lambda}_{\mbox{out}}(i)=\lambda_{\mbox{out}}$ for
$1\leq i \leq m$, we can have \be
P_{\mbox{out}}=P_{\mbox{out}}^{(1)}.\ee

\subsection{Another BER constraint condition}

In order to provide convenience for the system design, we hope
that $\lambda_0(i)=\overline{\lambda}_{\mbox{\small out}}(i)$. For
this reason, we revisit the derivation process of optimum solution
in Subsection II.B, and rewrite the expression (\ref{pp-2}) as \be
\label{pp-3} p_i=\left\{\begin{array}{ll}
\frac{\widehat{\mbox{SNR}}(i)}{\lambda_i}+\frac{2}{\beta_i\lambda_i}
\ln(\frac{\lambda_i\Delta(i)}{\overline{\lambda}_{\mbox{\small out}}(i)}), & \mbox{for} \;\lambda_i > \overline{\lambda}_{\mbox{\small out}}(i)\\
0, & \mbox{for} \; \lambda_i \leq \overline{\lambda}_{\mbox{\small
out}}(i)
     \end{array}
     \right.\ee
where the unknown optimization parameter $\Delta(i)$ should meet
the following BER constraint condition: \be  \label{ll-1}
\int_{\overline{\lambda}_{\mbox{\small out}}(i)}^{\infty}
\frac{\xi_i}{2}e^{-\beta_ip_i\lambda_i/2}f_i(\lambda_i)d\lambda_i=\overline{P}_o(i)\overline{P}_b(i).\ee
Substituting (\ref{pp-3}) and (\ref{Pb-1}) into (\ref{ll-1}), we
have after a simplifying process \be  \label{ll-2}
\overline{P}_b(i)\int_{\overline{\lambda}_{\mbox{\small
out}}(i)}^{\infty} \frac{\overline{\lambda}_{\mbox{\small
out}}(i)}{\lambda_i\Delta(i)}
f_i(\lambda_i)d\lambda_i=\overline{P}_o(i)\overline{P}_b(i).\ee
(\ref{ll-2}) can be simplified further to \be
\Delta(i)=\overline{\lambda}_{\mbox{\small out}}(i)
\int_{\overline{\lambda}_{\mbox{\small out}}(i)}^{\infty}
\frac{1}{\lambda_i} f_i(\lambda_i)d\lambda_i/\overline{P}_o(i).
\ee From the theorem of integral mean value, there is a constant
$g$ satisfying
 \be \int_{\overline{\lambda}_{\mbox{\small
out}}(i)}^{\infty} \frac{1}{\lambda_i}
f_i(\lambda_i)d\lambda_i=g\overline{P}_o(i). \ee  With $g$, we can
define a new function of $\overline{\lambda}_{\mbox{\small
 out}}(i)$ as follows:
 \be \label{mea} \overline{\lambda}_{\mbox{\small
 mea}}(i)=\frac{1}{g}=\frac{\overline{P}_o(i)}{\int_{\overline{\lambda}_{\mbox{\small out}}(i)}^{\infty}
\frac{1}{\lambda_i} f_i(\lambda_i)d\lambda_i} . \ee Furthermore,
it can follow from the theorem of integral mean value that \be
\overline{\lambda}_{\mbox{\small
 mea}}(i) \geq
\overline{\lambda}_{\mbox{\small out}}(i).\ee  Then $\Delta(i)$
can be rewritten as \be  \label{DD-1}
\Delta(i)=\frac{\overline{\lambda}_{\mbox{\small
out}}(i)}{\overline{\lambda}_{\mbox{\small
 mea}}(i)}.
\ee Taking account of the requirement of $p_i \geq 0$, from
(\ref{pp-3}) $\Delta(i)$ should also meet another BER constraint
condition: \be
\frac{\widehat{\mbox{SNR}}(i)}{\lambda_i}+\frac{2}{\beta_i\lambda_i}
\ln(\frac{\lambda_i\Delta(i)}{\overline{\lambda}_{\mbox{\small
out}}(i)})\geq 0 \ee or \be \Delta(i) \geq
\frac{\overline{\lambda}_{\mbox{\small
out}}(i)}{\lambda_i}e^{-\beta_i\widehat{\mbox{SNR}}(i)/2} \ee Due
to the fact of  $\lambda_i \geq \overline{\lambda}_{\mbox{\small
out}}(i)$, the constraint condition becomes under the help of
(\ref{Pb-1}) \be  \Delta(i) \geq \frac{2}{\xi_i}
\overline{P}_b(i). \ee By (\ref{DD-1}), the constraint condition
can be rewritten as \be \label{DD-2} \overline{P}_b(i) \leq
\frac{\xi_i}{2}\frac{\overline{\lambda}_{\mbox{\small
out}}(i)}{\overline{\lambda}_{\mbox{\small
 mea}}(i)}. \ee So we have the following lemma finally.
\begin{Lemma} If $\overline{P}_b(i) \leq
\frac{\xi_i}{2}\frac{\overline{\lambda}_{\mbox{\small
out}}(i)}{\overline{\lambda}_{\mbox{\small
 mea}}(i)}$, then,
\be \lambda_0(i)=\overline{\lambda}_{\mbox{\small out}}(i)\ee and
the optimum solution of power allocation is (\ref{pp-3}).
\end{Lemma}

\subsection{Minimum average transmit power}

Under the constraint (\ref{DD-2}), we now consider to derive the
minimum average transmit power. The derivation is not difficult,
but involves a process employing an unusual special function
appeared in \cite{Alouini}. Finally, we obtain the following
result and provide a detailed proof in Appendix.
\begin{Proposition} Suppose that $\overline{P}_b(i) \leq
\frac{\xi_i}{2}\frac{\overline{\lambda}_{\mbox{\small
out}}(i)}{\overline{\lambda}_{\mbox{\small
 mea}}(i)}$.  Let $\hat{\rho}^{(i)}(\overline{P}_b(i),P_{\mbox{out}}^{(i)})$ denote the average
needed transmit power for $i$-th data stream achieving the BER
given by (\ref{Pb-1}) under the condition that the OP
$P_{\mbox{out}}^{(i)}$ is given. Then \bee
\hat{\rho}^{(i)}(\overline{P}_b(i),P_{\mbox{out}}^{(i)})&=&
\rho_s(\overline{P}_b(i),P_{\mbox{out}}^{(i)})+\rho_{\Delta}(P_{\mbox{out}}^{(i)})\nnb \\
&=&\rho_s(\widehat{\mbox{SNR}}(i),\overline{\lambda}_{\mbox{\small
out}}(i))+\rho_{\Delta}(\overline{\lambda}_{\mbox{\small out}}(i))
\eee where \be \label{rr-5}
\rho_s(\widehat{\mbox{SNR}}(i),\overline{\lambda}_{\mbox{\small
out}}(i))=\sum_{k=i}^m(-1)^{k-i}{k-1 \choose i-1}{m \choose
k}\overline{F}_{\mbox{min:}k}^{\mbox{(pow)}}(\widehat{\mbox{SNR}}(i),\overline{\lambda}_{\mbox{\small
out}}(i)).\ee Moreover,
$\overline{F}_{\mbox{min:}k}^{\mbox{(pow)}}(\widehat{\mbox{SNR}}(i),\overline{\lambda}_{\mbox{\small
out}}(i))$ is the complementary distribution function of the
smallest random variable considered in a subset of $k$ random
variables over the random variable set of all eigenvalues, which
corresponds to
$\overline{F}_{\mbox{min:}k}^{\mbox{(out)}}(\overline{\lambda}_{\mbox{\small
out}}(i))$, and is given by \bee
&&\overline{F}_{\mbox{min:}k}^{\mbox{(pow)}}(\widehat{\mbox{SNR}}(i),\overline{\lambda}_{\mbox{\small
out}}(i))=\widehat{\mbox{SNR}}(i)
\cdot\frac{kC}{m!}\sum_{\mathcal{\alpha}}\sum_{\mathcal{\mu}}
\mbox{sgn}(\mathcal{\alpha})\mbox{sgn}(\mathcal{\mu})\times\nnb \\
&& A_{k}(\mathcal{\alpha},\mathcal{\mu})
\mathbb{\sum}_{\mathcal{\tau}}
\frac{\Gamma(\theta+\alpha_k+\mu_k-2+\sum_{\iota=1}^{k-1}\tau_{\iota},
k\overline{\lambda}_{\mbox{\small
out}}(i))}{k^{\theta+\alpha_k+\mu_k-2+\sum_{\iota=1}^{k-1}\tau_{\iota}}}
\nnb \\
&&\times\prod_{\iota=1}^{k-1}\frac{(\theta+\alpha_{\iota}+\mu_{\iota}-2)!}{\tau_{\iota}!}.
\eee

In the above equation, $\Gamma(q,x)$ is just the complementary
incomplete gamma function  (See Page 454 of \cite{Andrews}).

For the second term $\rho_{\Delta}(\cdot)$, it is independent of
$\widehat{\mbox{SNR}}(i)$, and is given by \be \label{delta}
\rho_{\Delta}(\overline{\lambda}_{\mbox{\small
out}}(i))=\sum_{k=i}^m(-1)^{k-i}{k-1 \choose i-1}{m \choose
k}\overline{F}_{\mbox{min:}k}^{\mbox{(del)}}(\overline{\lambda}_{\mbox{\small
out}}(i))\ee where
$\overline{F}_{\mbox{min:}k}^{\mbox{(del)}}(\overline{\lambda}_{\mbox{\small
out}}(i))$ corresponds to
$\overline{F}_{\mbox{min:}k}^{\mbox{(pow)}}(\widehat{\mbox{SNR}}(i),\overline{\lambda}_{\mbox{\small
out}}(i))$, and is given by \bee
&&\overline{F}_{\mbox{min:}k}^{\mbox{(del)}}(\overline{\lambda}_{\mbox{\small
out}}(i))=\frac{kC}{m!}\sum_{\mathcal{\alpha}}\sum_{\mathcal{\mu}}
\mbox{sgn}(\mathcal{\alpha})\mbox{sgn}(\mathcal{\mu})A_{k}(\mathcal{\alpha},\mathcal{\mu})\nnb \\
&& \times \mathbb{\sum}_{\mathcal{\tau}}
\frac{2\cdot\boldsymbol{\jmath}(N_k(\boldsymbol{\tau}),
k\overline{\lambda}_{\mbox{\small
out}}(i),\Delta(i))}{\beta_ik^{N_k(\boldsymbol{\tau})}}
\nnb \\
&&\times\prod_{\iota=1}^{k-1}\frac{(\theta+\alpha_{\iota}+\mu_{\iota}-2)!}{\tau_{\iota}!}
\eee with \be
N_k(\boldsymbol{\tau})=\theta+\alpha_k+\mu_k-2+\sum_{\iota=1}^{k-1}\tau_{\iota}
 \ee and \bee
&&\boldsymbol{\jmath}(N_k(\boldsymbol{\tau}),k\overline{\lambda}_{\mbox{\small
out}}(i),\Delta(i))=(N_k(\boldsymbol{\tau})-1)!
\cdot\sum_{\iota=0}^{N_k(\boldsymbol{\tau})-1}\frac{\Gamma(\iota,k\overline{\lambda}_{\mbox{\small
out}}(i))} {\iota!}\nnb \\
&&+\Gamma (N_k(\boldsymbol{\tau}),
k\overline{\lambda}_{\mbox{\small out}}(i))\cdot\ln \Delta(i).\eee
\end{Proposition}

\section{ Modified Power Allocation Scheme}

\subsection{A dynamic upper bound of Q function}
\begin{figure}[t]
\centering
\includegraphics[width=6.5 in]{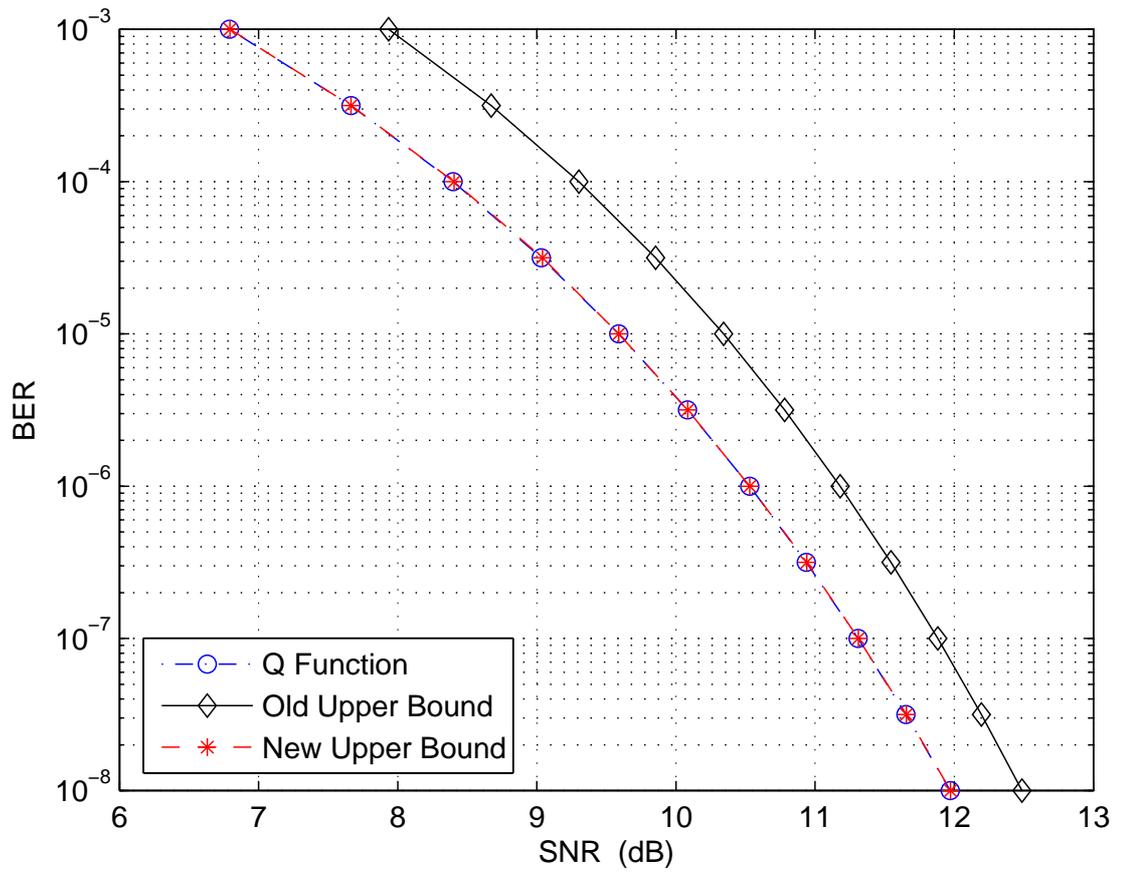}
\caption{BER comparison among Q function and its two upper bounds
for different SNRs} \label{fig.1}
\end{figure}

Fig.\ref{fig.1} makes comparison between Q function
$Q(\sqrt{2\mbox{SNR}})$ and its traditional upper bound
$\frac{1}{2}e^{-\mbox{SNR}/2}$. As seen in Fig.\ref{fig.1}, at the
important BER region their SNR deviation is relatively large and
slowly becomes small as SNR increases.  For example, when
$P_b=10^{-6}$, the SNR deviation is $0.65$ dB.

Therefore, in order to improve the system performance, we consider
to find a new upper bound of Q function replacing the old one and
with it give a modified power allocation scheme.  In order to
continue to employ the analysis method given in Section III, we
now need to study the following type of exponential upper bounds
of Q function: \be Q(\sqrt{2\mbox{SNR}}) \leq
\frac{1}{c(\mbox{SNR})} e^{-\mbox{SNR}/2}\ee Note that different
from the old upper bound, here we allow the designated parameter
c(SNR) to be dynamically variable.

For any given $\mbox{BER}=Q(\sqrt{2\mbox{SNR}})$, we easily find a
 $c(\mbox{SNR})$ which makes the new upper bound to approximate
 appropriately to the given BER. At the BER region from $10^{-3}$ to $10^{-8}$
 Fig.\ref{fig.1} also plots the new upper bound by using some appropriate values of $c(\mbox{SNR})$.

 As seen in Fig.\ref{fig.1}, the dynamic upper bound
approximates well to the exact value of Q function. The computed
results are also presented in Table I. From the table, it can be
observed that the optimized value of parameter $c$ increases as
the SNR increases. Therefore, we have the following property of Q
function:

\begin{Lemma} If $ Q(\sqrt{2\mbox{SNR}}) \leq \frac{1}{c(\mbox{SNR})}e^{-\mbox{SNR}/2} $ for a given SNR, then \be
Q(\sqrt{2(\mbox{SNR}+\Lambda)}) \leq \frac{1}{c(\mbox{SNR})}
e^{-(\mbox{SNR}+\Lambda)/2},\; \Lambda>0. \ee
\end{Lemma}

\subsection{Modified minimum ATP}

For any given $\overline{P}_b(i)$, we can find appropriate
$\widetilde{\mbox{SNR}}(i)$ and $c(\widetilde{\mbox{SNR}}(i))$
satisfying \be \label{Pb-2} \overline{P}_b(i)\approx
\frac{\xi_i}{c(\widetilde{\mbox{SNR}}(i))}e^{-\beta_i(i)\widetilde{\mbox{SNR}}(i)/2}.
\ee With the help of Lemma 3, thus the power allocation scheme can
be modified as
 \be \label{pp-4} p_i=\left\{\begin{array}{ll}
\frac{\widetilde{\mbox{SNR}}(i)}{\lambda_i}+\frac{2}{\beta_i\lambda_i}
\ln(\frac{\lambda_i\Delta(i)}{\overline{\lambda}_{\mbox{\small
out}}(i)}), & \mbox{for} \;\lambda_i > \overline{\lambda}_{\mbox{\small mea}}(i)\\
\frac{\widehat{\mbox{SNR}}(i)}{\lambda_i}+\frac{2}{\beta_i\lambda_i}
\ln(\frac{\lambda_i\Delta(i)}{\overline{\lambda}_{\mbox{\small
out}}(i)}), & \mbox{for} \;
\overline{\lambda}_{\mbox{\small out}}(i) <\lambda_i \leq \overline{\lambda}_{\mbox{\small mea}}(i)\\
0, & \mbox{for} \; \lambda_i \leq \overline{\lambda}_{\mbox{\small
out}}(i)\end{array} \right.\ee

With the modified power allocation scheme, the BER constraint
condition in (\ref{optimal-p}) can be still met since

\bee && \mathbb{E}P_b^{(i)} =
\int_{\overline{\lambda}_{\mbox{\small out}}(i)}^{\infty}
\xi_iQ(\sqrt{\beta_i\mbox{SNR}_i}) f_i(\lambda_i)d\lambda_i \nnb
\\&&= \int_{\overline{\lambda}_{\mbox{\small out}}(i)}^{\infty}
\xi_iQ(\sqrt{\beta_i p_i \lambda_i})
f_i(\lambda_i)d\lambda_i \nnb \\
&&= \int_{\overline{\lambda}_{\mbox{\small mea}}(i)}^{\infty}
\xi_iQ(\sqrt{\beta_i p_i \lambda_i}) f_i(\lambda_i)d\lambda_i
+\int_{\overline{\lambda}_{\mbox{\small
out}}(i)}^{\overline{\lambda}_{\mbox{\small
mea}}(i)}\xi_iQ(\sqrt{\beta_i p_i \lambda_i}) f_i(\lambda_i) d
\lambda_i \nnb \\
&&\leq \int_{\overline{\lambda}_{\mbox{\small mea}}(i)}^{\infty}
\frac{\xi_i}{c(\widetilde{\mbox{SNR}}(i))}e^{-\beta_ip_i\lambda_i/2}f_i(\lambda_i)d\lambda_i
+\int_{\overline{\lambda}_{\mbox{\small
out}}(i)}^{\overline{\lambda}_{\mbox{\small mea}}(i)}
\frac{\xi_i}{2}e^{-\beta_ip_i\lambda_i/2}f_i(\lambda_i)d\lambda_i
\nnb \\
&&\approx \overline{P}_b(i)\int_{\overline{\lambda}_{\mbox{\small
mea}}(i)}^{\infty} \frac{\overline{\lambda}_{\mbox{\small
out}}(i)}{\lambda_i\Delta(i)}
f_i(\lambda_i)d\lambda_i+\overline{P}_b(i)\int_{\overline{\lambda}_{\mbox{\small
out}}(i)}^{\overline{\lambda}_{\mbox{\small mea}}(i)}
\frac{\overline{\lambda}_{\mbox{\small
out}}(i)}{\lambda_i\Delta(i)} f_i(\lambda_i)d\lambda_i \nnb\\
&&=\overline{P}_b(i)\int_{\overline{\lambda}_{\mbox{\small
mea}}(i)}^{\infty} \frac{\overline{\lambda}_{\mbox{\small
out}}(i)}{\lambda_i\Delta(i)} f_i(\lambda_i)d\lambda_i
+\overline{P}_b(i)\int_{\overline{\lambda}_{\mbox{\small
out}}(i)}^{\overline{\lambda}_{\mbox{\small mea}}(i)}
\frac{\overline{\lambda}_{\mbox{\small
out}}(i)}{\lambda_i\Delta(i)} f_i(\lambda_i)d\lambda_i\nnb\\
&&=\overline{P}_b(i)\overline{P}_o(i). \eee

Moreover, we can verify  the fact that if (\ref{DD-2}) holds, the
optimal solution (\ref{pp-4}) exits.

Accordingly, the ATP for $i$-th subchannel is derived again and
Proposition 1  is modified as follows:

\begin{Proposition} Suppose that $\overline{P}_b(i) \leq
\frac{\xi_i}{2}\frac{\overline{\lambda}_{\mbox{\small
out}}(i)}{\overline{\lambda}_{\mbox{\small
 mea}}(i)}$.  Let $\tilde{\rho}^{(i)}(\overline{P}_b(i),P_{\mbox{out}}^{(i)})$ denote the average
needed transmit power for $i$-th data stream achieving the BER
given by (\ref{Pb-2}) under the condition that the OP
$P_{\mbox{out}}^{(i)}$ is given. Then \bee
\tilde{\rho}^{(i)}(\overline{P}_b(i),P_{\mbox{out}}^{(i)})&=&
\hat{\rho}^{(i)}(\overline{P}_b(i),P_{\mbox{out}}^{(i)})-\rho_s(\widehat{\mbox{SNR}}(i)-\widetilde{\mbox{SNR}}(i),\overline{\lambda}_{\mbox{\small
mea}}(i))
\nnb \\
&=&\rho_s(\widehat{\mbox{SNR}}(i),\overline{\lambda}_{\mbox{\small
out}}(i))-\rho_s(\widehat{\mbox{SNR}}(i)-\widetilde{\mbox{SNR}}(i),\overline{\lambda}_{\mbox{\small
mea}}(i))+\rho_{\Delta}(\overline{\lambda}_{\mbox{\small
out}}(i)). \eee

\end{Proposition}

\section{Numerical Results and Comparison}
In all of our simulation, we always make use of BPSK modulation
for each data stream transmission, which corresponds to the
modulation parameters $\xi_i=1, \beta_i=2, i=1,2,\ldots, m$. For
simplicity, we let each data stream have the same constraint
parameters, i.e., \be
P_{\mbox{out}}^{(1)}=P_{\mbox{out}}^{(2)}=\ldots=P_{\mbox{out}}^{(i)}=P_{\mbox{out}},
\ee and \be
\overline{P}_b(1)=\overline{P}_b(2)=\ldots=\overline{P}_b(m)=P_b.
\ee We first observe the behavior of MIMO individual outage
probability using Lemma 1. In order to provide convenience for
making OP comparison between SISO and MIMO systems, we first
evaluate outage probability $P_{\mbox{out}}=10^{-\upsilon}$ for
SISO systems by setting exponent $\upsilon$.  And we call
$\upsilon$ as a SISO outage exponent (OE). For example,  we set
the OE $\upsilon=1$, then $P_{\mbox{out}}=10^{-1}$ and $
\lambda_{\mbox{out}}=1.1 \cdot 10^{-1}$ for the SISO system, and
under given $ \lambda_{\mbox{out}}=1.1 \cdot 10^{-1}$ we can
compute further $P_{\mbox{out}}=6.6 \cdot 10^{-5}$ for the MIMO
system with $n=6$, $m=3$, and $i=3$. Table II provides computed
results for the MIMO system with $n=6$, $m=3$, and $i=3$ when
$\upsilon$ is set from $0.4$ to $1.8$. Table II shows that the
MIMO system has lower individual OP as $\upsilon$ increases, and
almost has no outage when $\upsilon \geq 1.6$. In the following,
if needed, we will always set $\lambda_{\mbox{out}}=1.1 \cdot
10^{-1}$, whose corresponding OE is $\upsilon=1$ .

\begin{figure}[t]
\centering
\includegraphics[width=6.5 in]{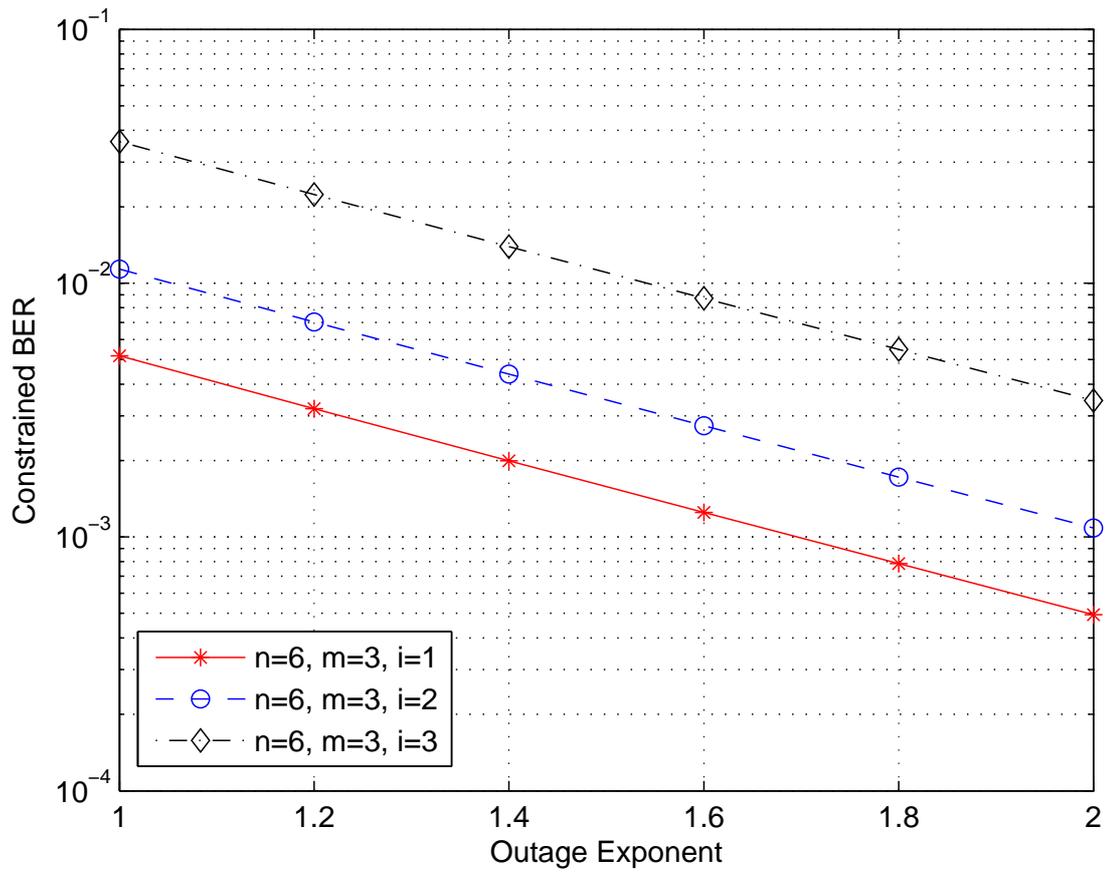}
\caption{Constrained BER under various SISO outage exponents}
\label{fig.2}
\end{figure}

We now consider the constraint condition of BER in optimization
design which is given in (\ref{DD-2}). Fig.\ref{fig.2} plots the
constrained BER for the MIMO system with $m=3$ and $n=6$ when the
OE is set appropriately from $1$ to $2$. From this figure, the
constraint condition is easily met for any of the three data
streams $i=1, 2, 3$.

\begin{figure}[t]
\centering
\includegraphics[width=6.5 in]{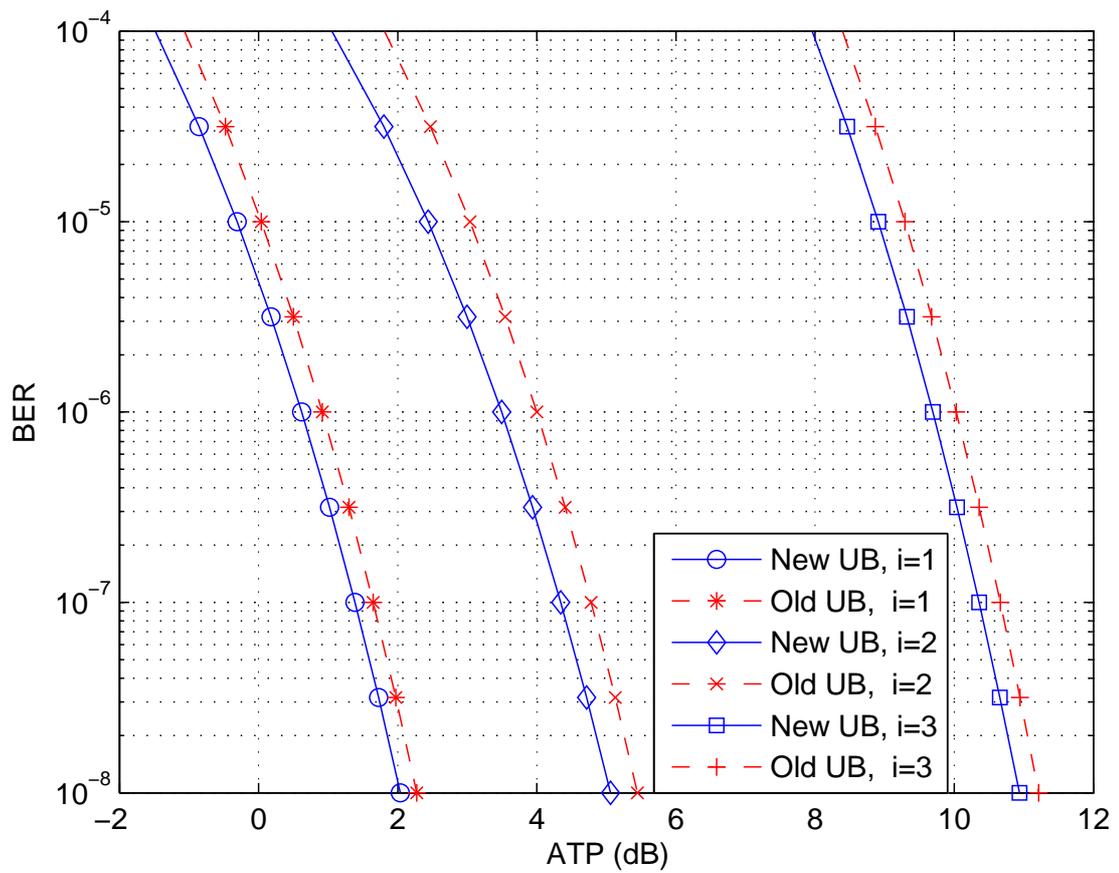}
\caption{ATP comparison between two power allocation schemes under
different average BER constraints} \label{fig.3}
\end{figure}

We still fix the minimum antenna number $m=3$ and the maximum
antenna number $n=6$. Fig.\ref{fig.3} plots the individual ATP for
the two adaptive transmit schemes produced using the new and old
upper bounds of Q function for $i=1,  2,  3$. It can be observed
that the ATP increases gradually as $i$ increases, which implies
that the channel condition becomes worse. On the other hand, the
power allocation with the new upper bound (UB) has more power
saving than the one with the old UB for all $i=1,  2,  3$.

\begin{figure}[t]
\centering
\includegraphics[width=6.5 in]{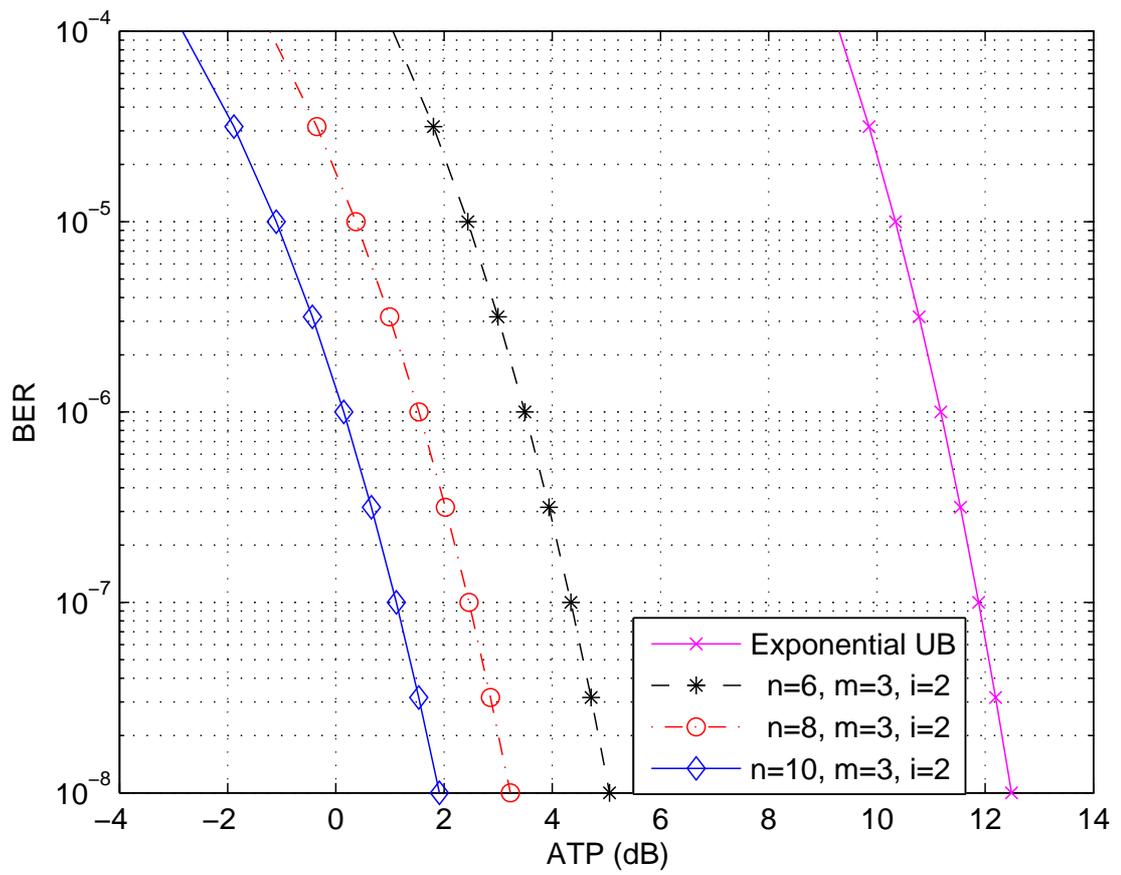}
\caption{ATP against the average BER for different $n$}
\label{fig.4}
\end{figure}

Finally, we fix $m=3$ and $i=2$.  Fig.\ref{fig.4} plots the
individual ATP computed by Proposition 2 for different $n$. It can
be observed that as the maximum number of antennas $n$ increases,
the needed ATP decreases, but the amount of ATP improvement
becomes gradually small. For comparison, this figure also includes
the BER curve with the traditional UB of Q function
$\mbox{BER}=\frac{1}{2}e^{-\widehat{\mbox{SNR}}}$.  It can be
found from this figure that all of the three different MIMO
configurations have the reliable performance of exponential
average BER.

\section{Conclusions}

It is well known that multiple antennas can provide high
multiplexing and diversity gains for wireless communications.
Adaptive transmission techniques in wireless communications can
utilize the system resources efficiently and provide satisfactory
QoS. In this paper, we have investigated adaptive transmission
technique mainly based on channel eigenvalues for MIMO multi-beams
systems. Under the BER and OP constraints, we have presented the
closed-form expressions for the minimum average transmit power and
individual outage probability.  Our theoretical analysis further
shows that in fading environments wireless communications
employing multiple antennas can also achieve the exponential BER
performance, as in AWGN channels.

\appendix
{\em The proof of Proposition 1:} Since $\overline{P}_b(i) \leq
\frac{\xi_i}{2}\frac{\overline{\lambda}_{\mbox{\small
out}}(i)}{\overline{\lambda}_{\mbox{\small
 mea}}(i)}$, then it follows from Lemma 2 that
$\lambda_0(i)=\overline{\lambda}_{\mbox{\small out}}(i) $.
Furthermore, we can have \be \label{app1}
\hat{\rho}^{(i)}(\overline{P}_b(i),P_{\mbox{out}}^{(i)})=\int_{\overline{\lambda}_{\mbox{\small
out}}(i)}^{\infty}p_i f_i(\lambda_i)d\lambda_i.\ee Substituting
(\ref{pp-3}) into (\ref{app1}), we have \be
\hat{\rho}^{(i)}(\overline{P}_b(i),P_{\mbox{out}}^{(i)})=\int_{\overline{\lambda}_{\mbox{\small
out}}(i)}^{\infty}\frac{\widehat{\mbox{SNR}}(i)}{\lambda_i}
f_i(\lambda_i)d\lambda_i+\int_{\overline{\lambda}_{\mbox{\small
out}}(i)}^{\infty}\frac{2}{\beta_i\lambda_i}
\ln(\frac{\lambda_i\Delta(i)}{\overline{\lambda}_{\mbox{\small
out}}(i)}) f_i(\lambda_i)d\lambda_i.\ee  Define \be \label{app2}
\rho_s(\widehat{\mbox{SNR}}(i),\overline{\lambda}_{\mbox{\small
out}}(i))=\int_{\overline{\lambda}_{\mbox{\small
out}}(i)}^{\infty}\frac{\widehat{\mbox{SNR}}(i)}{\lambda_i}
f_i(\lambda_i)d\lambda_i   \ee and \be \label{app3}
\rho_{\Delta}(\overline{\lambda}_{\mbox{\small
out}}(i))=\int_{\overline{\lambda}_{\mbox{\small
out}}(i)}^{\infty}\frac{2}{\beta_i\lambda_i}
\ln(\frac{\lambda_i\Delta(i)}{\overline{\lambda}_{\mbox{\small
out}}(i)}) f_i(\lambda_i)d\lambda_i. \ee

Now we consider to derive (\ref{app2}) and (\ref{app3}),
respectively. From Lemma 1 in \cite{Zanella}, the marginal p.d.f.
of the $i$-th largest eigenvalue $\lambda_i$ can be written as \be
\label{app4} f_i(\lambda_i)=\sum_{k=i}^m(-1)^{k-i}{k-1 \choose
i-1}{m \choose k}f_{\mbox{min:}k}(\lambda_i)\ee where
$f_{\mbox{min:}k}(x)$ denotes the p.d.f. of the smallest random
variable considered in a subset of $k$ random variables over the
set of all eigenvalues, and is given by \bee
&&f_{\mbox{min:}k}(x)=\frac{kC}{m!}\sum_{\mathcal{\alpha}}\sum_{\mathcal{\mu}}
\mbox{sgn}(\mathcal{\alpha})\mbox{sgn}(\mathcal{\mu})A_{k}(\mathcal{\alpha},\mathcal{\mu})\nnb \\
&& \times e^{-kx}\mathbb{\sum}_{\mathcal{\tau}}
x^{\theta+\alpha_k+\mu_k-2+\sum_{\iota=1}^{k-1}\tau_{\iota}}
\nnb \\
&&\times\prod_{\iota=1}^{k-1}\frac{(\theta+\alpha_{\iota}+\mu_{\iota}-2)!}{\tau_{\iota}!}.
\eee  So with the help of the complementary incomplete gamma
function $\Gamma(q,x)$, we can obtain the desired result
(\ref{rr-5}) after a simple derivation.

The derivation of (\ref{app3}) is similar, but involves a process
employing the following special function $\jmath_q(x)$ defined as
\cite{Alouini}: \bee  \jmath_q(x)&=& \int_1^\infty t^{q-1}\ln t e^{-xt}dt\nnb\\
&=& \frac{(q-1)!}{x^q}\sum_{k=0}^{q-1}\Gamma(k,x)/k!.\eee Finally,
again making use of (\ref{app4}), we can easily obtain the desired
expression of $\rho_{\Delta}(\overline{\lambda}_{\mbox{\small
out}}(i))$ (\ref{delta}).

\balance

\bibliographystyle{ieeetr}

\newpage

\begin{table*}[htb!]
\begin{center}
\caption{ Minimization parameter $c$ for the new upper bound of Q
function}
\begin{tabular}{|l|r|r|r|r|r|r|r|r|r|r|r|}\hline\hline
  BER   & $10^{-3}$ & $10^{-3.5}$ & $10^{-4}$ & $10^{-4.5}$
 & $10^{-5}$& $10^{-5.5}$& $10^{-6}$& $10^{-6.5}$& $10^{-7}$ &$10^{-7.5}$ & $10^{-8}$ \\\hline
 SNR     &6.8   &7.7   &8.4  &9.0  &9.6   &10.1  &10.5 &10.9 & 11.3 & 11.7 & 12.0  \\\hline
 $c(\mbox{SNR})$ &8.4   &9.2   &9.8  &10.4  &11.2   &11.8  &12.4 &12.8 & 13.4 & 13.8 & 14.4 \\\hline\hline
\end{tabular}
\end{center}
\end{table*}

\begin{table*}[htb!]
\begin{center}
\caption{Outage probability comparison between SISO and MIMO
systems}
\begin{tabular}{|l|r|r|r|r|r|r|r|r|}\hline\hline
 SISO OE     &0.4   &0.6   &0.8  &1.0  &1.2   &1.4  &1.6   &1.8 \\\hline
$\overline{\lambda}_{\mbox{\small out}}$  & $5.1 \times 10^{-1}$ &
$2.9 \times 10^{-1}$& $1.7 \times 10^{-1}$& $1.1 \times
10^{-1}$&$6.5 \times 10^{-2}$&$4.1 \times 10^{-2}$&$2.5 \times
10^{-2}$&$1.6 \times 10^{-2}$\\\hline
    SISO OP & $4.0 \times 10^{-1}$&$
2.5 \times 10^{-1}$&$ 1.6 \times 10^{-1}$&$ 1.0 \times
10^{-1}$&$6.3 \times 10^{-2}$&$4.0 \times 10^{-2}$&$2.5 \times
10^{-2}$&$1.6 \times 10^{-2}$\\\hline
    MIMO OP & $2.0 \times 10^{-2}$&$
2.9 \times 10^{-3}$&$ 4.3 \times 10^{-4}$&$ 6.6 \times
10^{-5}$&$1.0 \times 10^{-5}$&$1.6 \times 10^{-6}$&$2.5 \times
10^{-7}$&$4.0 \times 10^{-8}$\\\hline\hline
\end{tabular}
\end{center}
\end{table*}

\end{document}